\documentclass[prl,showpacs,superscriptaddress,floatfix,twocolumn]{revtex4}
\usepackage{amsfonts}
\usepackage{stmaryrd}
\usepackage{bbm}
\usepackage{mathrsfs}
\usepackage{tipa}
\usepackage{amssymb}
\usepackage{txfonts}
\usepackage{graphicx}
\usepackage{dcolumn}
\usepackage{epstopdf}
\usepackage{array}

\newcommand{\PreserveBackslash}[1]{\let\temp=\\#1\let\\=\temp}
\newcolumntype{C}[1]{>{\PreserveBackslash\centering}p{#1}}
\newcolumntype{R}[1]{>{\PreserveBackslash\raggedleft}p{#1}}
\newcolumntype{L}[1]{>{\PreserveBackslash\raggedright}p{#1}}

\begin{document}

\newcommand*{\cm}{cm$^{-1}$\,}

\title{Compensated Semimetal LaSb with Unsaturated Magnetoresistance}

\author{L.-K. Zeng}\thanks{These authors contributed equally to this work.}
\affiliation{Beijing National Laboratory for Condensed Matter Physics, and Institute of Physics, Chinese Academy of Sciences, Beijing 100190, China}

\author{R. Lou}\thanks{These authors contributed equally to this work.}
\affiliation{Department of Physics and Beijing Key Laboratory of Opto-electronic Functional Materials $\textsl{\&}$ Micro-nano Devices, Renmin University of China, Beijing 100872, China}

\author{D.-S. Wu}\thanks{These authors contributed equally to this work.}
\affiliation{Beijing National Laboratory for Condensed Matter Physics, and Institute of Physics, Chinese Academy of Sciences, Beijing 100190, China}

\author{Q. N. Xu}
\affiliation{Beijing National Laboratory for Condensed Matter Physics, and Institute of Physics, Chinese Academy of Sciences, Beijing 100190, China}

\author{P.-J. Guo}
\affiliation{Department of Physics and Beijing Key Laboratory of Opto-electronic Functional Materials $\textsl{\&}$ Micro-nano Devices, Renmin University of China, Beijing 100872, China}

\author{L.-Y. Kong}
\author{Y.-G. Zhong}
\author{J.-Z. Ma}
\author{B.-B. Fu}
\affiliation{Beijing National Laboratory for Condensed Matter Physics, and Institute of Physics, Chinese Academy of Sciences, Beijing 100190, China}

\author{P. Richard}
\affiliation{Beijing National Laboratory for Condensed Matter Physics, and Institute of Physics, Chinese Academy of Sciences, Beijing 100190, China}
\affiliation{Collaborative Innovation Center of Quantum Matter, Beijing, China}

\author{P. Wang}
\affiliation{Beijing National Laboratory for Condensed Matter Physics, and Institute of Physics, Chinese Academy of Sciences, Beijing 100190, China}

\author{G. T. Liu}
\affiliation{Beijing National Laboratory for Condensed Matter Physics, and Institute of Physics, Chinese Academy of Sciences, Beijing 100190, China}

\author{L. Lu}
\affiliation{Beijing National Laboratory for Condensed Matter Physics, and Institute of Physics, Chinese Academy of Sciences, Beijing 100190, China}
\affiliation{Collaborative Innovation Center of Quantum Matter, Beijing, China}

\author{Y.-B. Huang}
\affiliation{Shanghai Synchrotron Radiation Facility, Shanghai Institute of Applied Physics, Chinese Academy of Sciences, Shanghai 201204, China}

\author{C. Fang}
\affiliation{Beijing National Laboratory for Condensed Matter Physics, and Institute of Physics, Chinese Academy of Sciences, Beijing 100190, China}

\author{S.-S. Sun}
\author{Q. Wang}
\affiliation{Department of Physics and Beijing Key Laboratory of Opto-electronic Functional Materials $\textsl{\&}$ Micro-nano Devices, Renmin University of China, Beijing 100872, China}

\author{L. Wang}
\author{Y.-G. Shi}
\affiliation{Beijing National Laboratory for Condensed Matter Physics, and Institute of Physics, Chinese Academy of Sciences, Beijing 100190, China}

\author{H. M. Weng}
\affiliation{Beijing National Laboratory for Condensed Matter Physics, and Institute of Physics, Chinese Academy of Sciences, Beijing 100190, China}
\affiliation{Collaborative Innovation Center of Quantum Matter, Beijing, China}

\author{H.-C. Lei}
\author{K. Liu}
\author{S.-C. Wang}
\email{scw@ruc.edu.cn}
\affiliation{Department of Physics and Beijing Key Laboratory of Opto-electronic Functional Materials $\textsl{\&}$ Micro-nano Devices, Renmin University of China, Beijing 100872, China}

\author{T. Qian}
\email{tqian@iphy.ac.cn}
\affiliation{Beijing National Laboratory for Condensed Matter Physics, and Institute of Physics, Chinese Academy of Sciences, Beijing 100190, China}
\affiliation{Collaborative Innovation Center of Quantum Matter, Beijing, China}

\author{J.-L. Luo}
\email{jlluo@iphy.ac.cn}
\affiliation{Beijing National Laboratory for Condensed Matter Physics, and Institute of Physics, Chinese Academy of Sciences, Beijing 100190, China}
\affiliation{Collaborative Innovation Center of Quantum Matter, Beijing, China}

\author{H. Ding}
\affiliation{Beijing National Laboratory for Condensed Matter Physics, and Institute of Physics, Chinese Academy of Sciences, Beijing 100190, China}
\affiliation{Collaborative Innovation Center of Quantum Matter, Beijing, China}

\begin{abstract}
  By combining angle-resolved photoemission spectroscopy and quantum oscillation measurements, we performed a comprehensive investigation
  on the electronic structure of LaSb, which exhibits near-quadratic extremely large magnetoresistance (XMR) without any sign of saturation
  at magnetic fields as high as 40 T. We clearly resolve one spherical and one intersecting-ellipsoidal hole Fermi surfaces (FSs) at the
  Brillouin zone (BZ) center $\Gamma$ and one ellipsoidal electron FS at the BZ boundary $X$. The hole and electron carriers calculated
  from the enclosed FS volumes are perfectly compensated, and the carrier compensation is unaffected by temperature. We further reveal
  that LaSb is topologically trivial but share many similarities with the Weyl semimetal TaAs family in the bulk electronic structure.
  Based on these results, we have examined the mechanisms that have been proposed so far to explain the near-quadratic XMR in semimetals.
\end{abstract}

\pacs{75.47.-m, 71.18.+y, 79.60.-i, 71.20.Eh}

\maketitle

Magnetoresistance (MR) is the magnetic-field induced changes of electrical resistance of a material, which has attracted great
attention not only in understanding the underlying physical mechanisms but also for practical applications, such as spintronics
devices, magnetic memory and magnetic field sensors. Negative MR has been discovered in many magnetic materials, such as giant
MR in magnetic multilayer films \cite{1, 2} and colossal MR in perovskite manganites \cite{3, 4}. Extremely large positive MR
(XMR) has been reported in nonmagnetic materials \cite{5, 6, 7, 8, 9, 10}.

Recently, the discovery of XMR without any sign of saturation at magnetic fields up to 60 T in nonmagnetic semimetal WTe$_2$ has renewed
the research interest for this topic \cite{11}. The magnetotransport in WTe$_2$ is characterized by a typical near-quadratic field dependence
of MR and a field-induced up-turn in resistivity followed by a plateau at low temperature. Soon after, these fingerprints were also observed
in several semimetals including $TmPn_2$ ($Tm$ = Ta/Nb, $Pn$ = As/Sb) \cite{12, 13, 14, 15, 16, 17}, $LnX$ ($Ln$ = La/Y, $X$ = Sb/Bi) \cite{
18, 19, 20, 21, 22}, and ZrSiS \cite{23, 24, 25}. These common features imply that the quadratic XMR in these nonmagnetic semimetals may have
the same origin. XMR has also been observed in Dirac and Weyl semimetals like Cd$_3$As$_2$ \cite{26, 27}, TaAs \cite{28}, and NbP \cite{29},
which, nevertheless, show a linear field dependence of MR distinct from the near-quadratic behavior. So far, several mechanisms have been
proposed to explain the exotic quadratic XMR behavior, including electron-hole resonance compensation \cite{11, 30}, forbidden backscattering
at zero field \cite{31}, field-induced Fermi surface (FS) changes \cite{12}, and nontrivial band topology \cite{18}.

It is widely believed that XMR in semimetals is intimately related to their underlying electronic structures. First-principles
calculations have shown that most of their electronic structures are rather complicated \cite{10, 11, 12, 13, 14, 15, 32}. The
experimental electronic structures by angle-resolved photoemission spectroscopy (ARPES) measurements are even more complicated
and illegible because of the convoluted bulk and surface states \cite{31, 33, 34, 35, 36, 37, 38}. The lack of unambiguous
experimental data on the intrinsic electronic structures seriously obstructs the investigation on the underlying mechanism
of quadratic XMR.

In this work, we report a comprehensive study on the electronic structure of LaSb by combining ARPES and quantum oscillation (QO)
measurements. The FS topology of LaSb is clearly resolved, which consists of two hole FS pockets at the Brillouin zone (BZ) center
$\Gamma$ and one electron FS pocket at the BZ boundary $X$. We have precisely quantified the hole and electron carrier densities
and demonstrated that LaSb is a compensated semimetal. The measured band structure confirms that LaSb is topologically trivial but
with a linearly dispersive bulk band. As compared to other XMR semimetals, LaSb has a much simpler electronic structure, which
facilitates the investigation on the origin of quadratic XMR.

High-quality single crystals of LaSb were grown by the flux method. ARPES measurements were performed at the Dreamline beamline of the
Shanghai Synchrotron Radiation Facility (SSRF) with a Scienta D80 analyzer and at the beamline 13U of the National Synchrotron Radiation
Laboratory (NSRL) at Hefei with a Scienta R4000 analyzer. The energy and angular resolutions were set to 15 meV and 0.05$^\circ$,
respectively. The samples were cleaved $\emph{in situ}$ along the (001) plane and measured at $T$ = 30 and 200 K in a working vacuum
better than 5$\times$10$^{-11}$ Torr.

\begin{figure}[htb]
  \begin{center}
  \includegraphics[width=1\columnwidth]{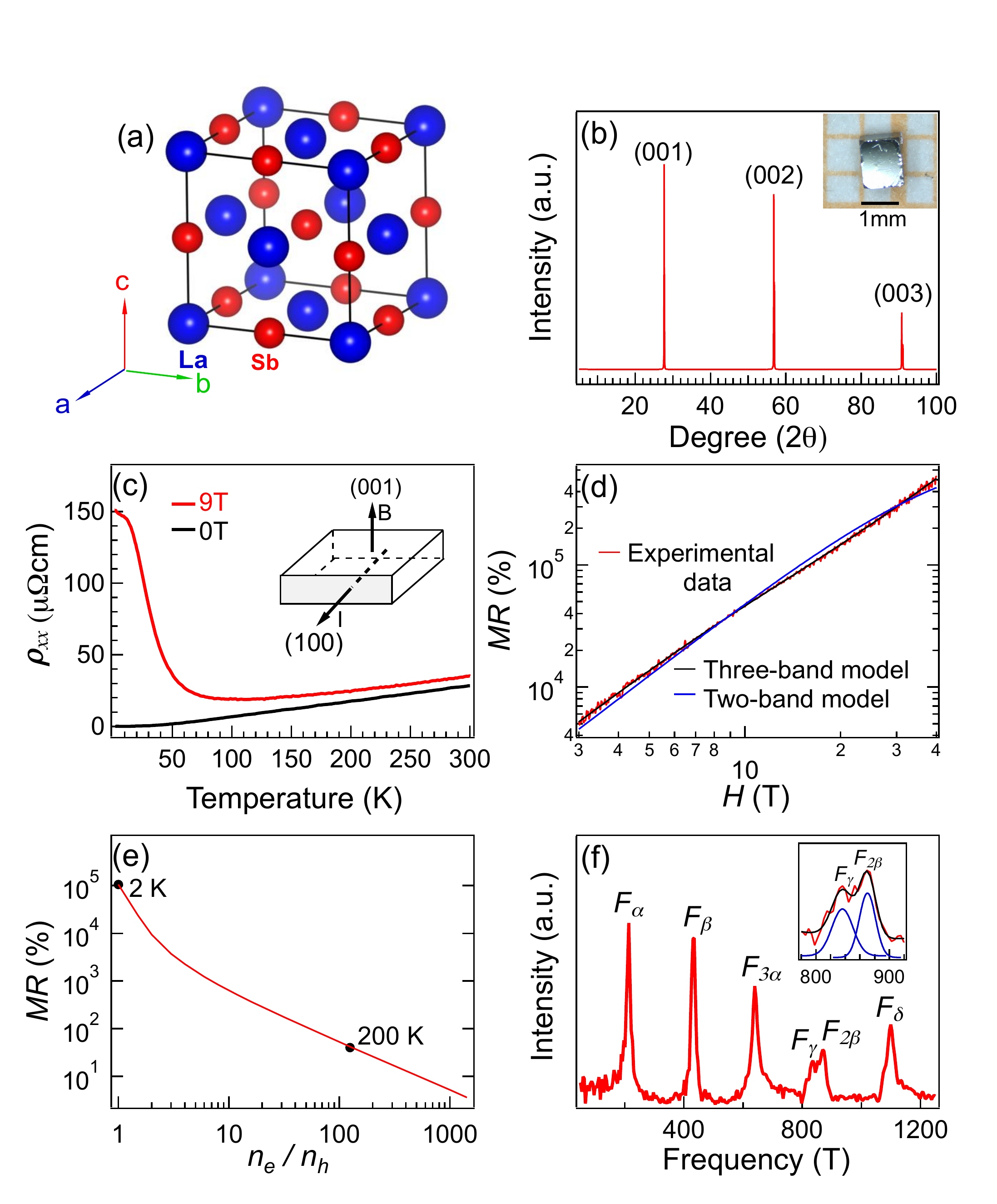}
  \end{center}
  \caption{(Color online) Crystal structure and magnetotransport of LaSb.
  (a) Schematic crystal structure of LaSb.
  (b) XRD pattern on the (001) surface of single crystal. The inset shows a typical cuboid single crystal.
  (c) Resistivity as a function of temperature at magnetic field $H$ = 0 (black curve) and 9 T (red curve). The inset illustrates the
      directions of $H$ and $I$ in the magnetotransport measurements.
  (d) $MR$ (\%) = [$R$($H$) - $R$(0)]/$R$(0) $\times$ 100 \% plotted as a function of field up to 40 T at $T$ = 2 K (red curve), where 
      $R$($H$) and $R$(0) represent the resistivity at magnetic field $H$ and at zero field, respectively. Blue curve is the fitting to 
      the two-band model considering slightly imperfect carrier compensation ($n_e$/$n_h$ = 0.998). Black curve is the fitting to the
      three-band model with perfect carrier compensation.
  (e) Simulated MR as a function of the ratio $n_e$/$n_h$ based on the two-band model. Solid circles represent the experimental MR at $T$
      = 2 and 200 K under 9 T assuming that the suppression of MR at high temperatures is attributed to carrier imbalance.
  (f) FFT spectrum of the Shubnikov-de Haas oscillations shown in the Supplemental Materials. The inset shows that the peak positions of
      $F_{\gamma}$ and $F_{2\beta}$ are extracted by the fitting to two Gaussian functions.}
\end{figure}

Figure 1(a) shows the schematic of the crystal structure of LaSb. It has a simple rock salt structure, which is face-center
cubic with space group $Fm$-3$m$. A typical cuboid single crystal of LaSb is shown in the inset of Fig. 1(b). X-ray diffraction
(XRD) measurements on single crystals confirm that the rectangular face is the (001) plane. The electric current ($I$) and the
magnetic field ($H$) in our magnetotransport measurements are applied along the (100) and (001) directions, respectively, as
illustrated in the inset of Fig. 1(c). In Fig. 1(c), the resistivity at zero field shows a metallic behavior in the measured
temperature range from 2 to 300 K. When a magnetic field of 9 T is applied, the resistivity shows a minimum at a field-induced
``turn on'' temperature $T$ $\sim$ 100 K, and then increases dramatically with decreasing temperature. A resistivity plateau
is ultimately formed after an inflection at $T$ $\sim$ 15 K, which is consistent with previous results on LaSb \cite{18}.

In Fig. 1(d), the MR curve as a function of magnetic field exhibits no any sign of saturation even up to 40 T. Unsaturated MR at
magnetic fields up to several tens of Tesla has also been reported in other XMR semimetals like WTe$_2$ \cite{11}, LaBi \cite{20},
and ZrSiS \cite{25}. The MR curve can be approximately fit to a power-law function with an exponent $m$ = 1.78. Similar power-law
field dependence with $m$ slightly less than 2 has been observed in WTe$_2$ \cite{39}, PtSn$_4$ \cite{10}, NbSb$_2$ \cite{12}, and
LaBi \cite{19}. So far, the most commonly used mechanism to explain the quadratic XMR is the two-band model, which predicts a
quadratic field dependence of MR as electron-hole compensation is satisfied \cite{40}. The deviation from the quadratic behavior
was simply attributed to slight imbalance between electrons and holes. We fit the data based on the two-band model considering
imperfect carrier compensation \cite{19, 41} and obtain the ratio of electron-to-hole carrier density $n_e$/$n_h$ = 0.998 by
assuming equal electron and hole mobilities. Although the deviation from perfect compensation is negligible, the fitting curve
exhibits an obvious trend of saturation, which does not reproduce the MR data. The two-band model is not sufficient to explain
the near-quadratic behavior in LaSb, and unlikely the deviation from quadratic in other XMR materials either. To understand the
XMR in LaSb, we investigate comprehensively its electronic structure.

\begin{figure}[htb]
  \begin{center}
  \includegraphics[width=1\columnwidth]{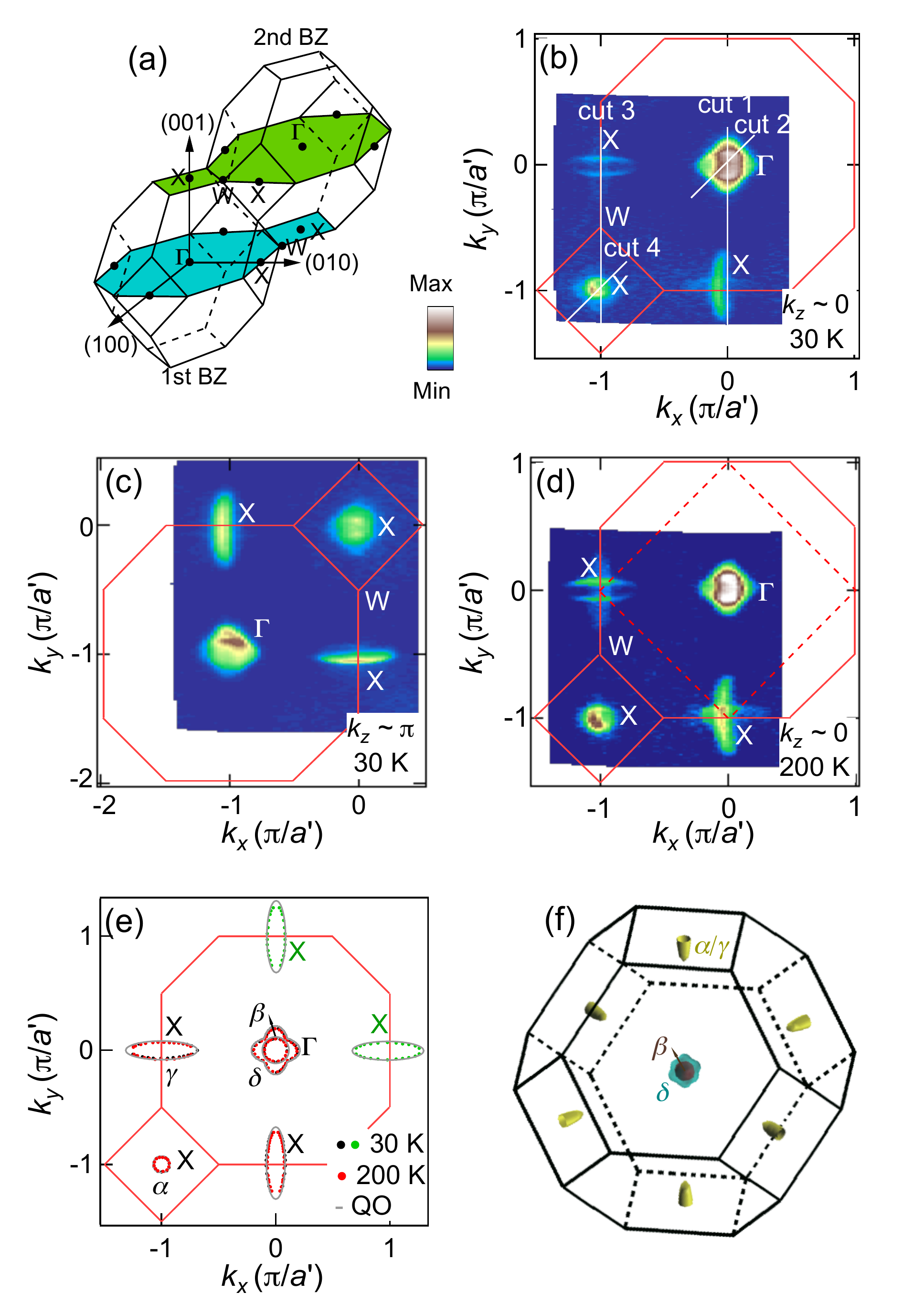}
  \end{center}
  \caption{(Color online) Experimental and calculated FSs of LaSb.
  (a) Schematic of the first and second 3D BZs. The cyan and green areas illustrate the locations of the mapping data in (b) and (c),
      respectively.
  (b),(c) FS intensity plots obtained by integrating the photoemission intensity within $E_F$$\pm$10 meV recorded with $hv$ = 53 and 83 eV,
          close to the $k_z$ = 0 and $\pi$ planes, respectively, at $T$ = 30 K. $a'$ is the half of lattice constant $a$ (= 6.5 $\AA$) of
          the face-center-cubic unit cell. Cuts 1-4 in (b) indicate the momentum locations of the measured bands in Fig. 3.
  (d) Same as (b), but measured at $T$ = 200 K. Dashed lines represent the (001) surface BZ.
  (e) Fermi wave vectors extracted from the data recorded with $hv$ = 53 (black dots) and 83 eV (green dots) at $T$ = 30 K and $hv$ = 53
      eV at $T$ = 200K (red dots). Solid lines are FSs derived from QOs.
  (f) Calculated 3D FSs with the MBJ potential.}
\end{figure}

\begin{figure*}[htb]
  \begin{center}
  \includegraphics[width=1.7\columnwidth]{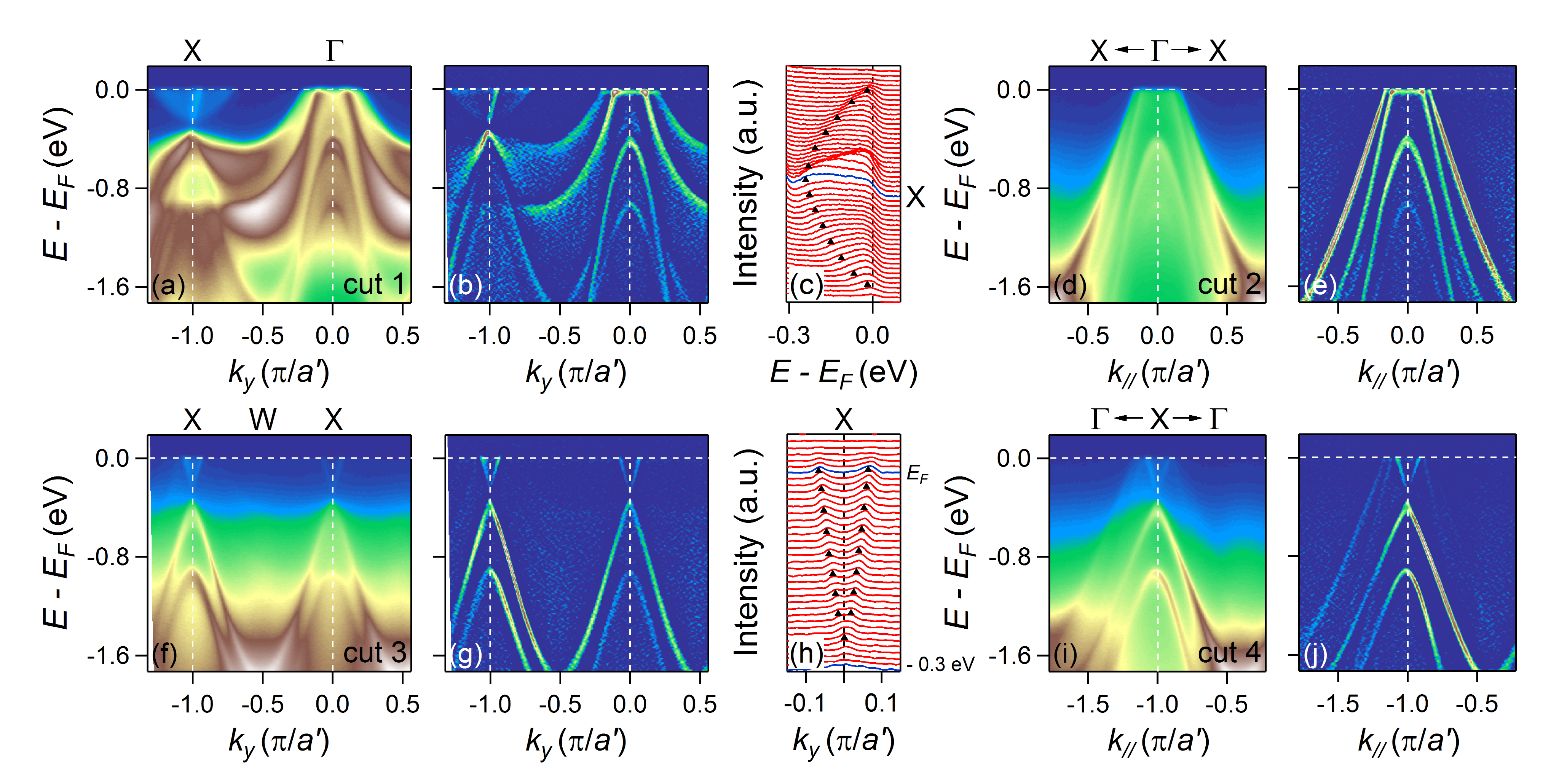}
  \end{center}
  \caption{(Color online) Near-$E_F$ band dispersions along high-symmetry lines measured at $k_z$ $\sim$ 0 plane ($hv$ = 53 eV).
  (a),(b) Photoemission intensity plot along $\Gamma$-$X$ [cut 1 in Fig. 2(b)] and corresponding 2D curvature intensity plot \cite{48},
          respectively.
  (c) Energy distribution curves of (a) around $X$.
  (d),(e) Same as (a),(b) but rotated by 45$^{\circ}$ [cut 2 in Fig. 2(b)].
  (f),(g) Same as (a),(b) but along $X$-$W$-$X$ [cut 3 in Fig. 2(b)].
  (h) Momentum distribution curves of (f) around the right $X$.
  (i),(j) Same direction as in (d),(e) but around $X$ (-$\pi$, -$\pi$) [cut 4 in Fig. 2(b)]. The black dots in (c) and (h) indicate the
          peak positions, which track the dispersions of the electron band along and perpendicular to the long-axis of the ellipsoidal
          $\gamma$ FS, respectively. The linear electron band at $X$ in (a)-(c) and the hole bands around $X$ in (i) and (j) are a result
          of the band folding with a wave vector $Q$ = ($\pi$, $\pi$, 0).}
\end{figure*}

We demonstrate that the electron and hole carriers in LaSb are perfectly compensated by combining QO and ARPES measurements. The fast
Fourier transform (FFT) spectrum of the QO data in Fig. 1(f) exhibits four principal frequencies: $F_\alpha$ = 214 T with its third
harmonic $F_{3\alpha}$ = 644 T, $F_\beta$ = 436 T with its second harmonic $F_{2\beta}$ = 875 T, $F_\gamma$ = 836 T, and $F_\delta$ =
1090 T. Using the Onsager relation [$F$ = ($\hbar$/$2{\pi}e$)$A_k$ between frequency $F$ and the extreme cross section $A_k$ of a FS],
we extract the corresponding $A_k$ values.

We further determine the topology of these FSs by ARPES measurements. Figures 2(b) and 2(c) show the intensity maps at the Fermi level
($E_F$) recorded at $T$ = 30 K with photon energy $hv$ = 53 and 83 eV, close to the $k_z$ = 0 and $\pi$ planes, respectively (See more
photon energy dependence data in the Supplemental Materials). To examine temperature effects on the electronic structure, we also performed
ARPES measurements with $hv$ = 53 eV at $T$ = 200 K and plot the FS intensity map in Fig. 2(d). The FS topology at $k_z$ = 0 [Fig. 2(b)] is
the same as at $k_z$ = $\pi$ [Fig. 2(c)], but shifted by the wave vector ($\pi$, $\pi$), which is illustrated in the schematic three-dimensional
(3D) BZ in Fig. 2(a). We extract the Fermi wave vectors of the FSs and plot them as symbols in Fig. 2(e). The FSs at $k_z$ = 0 in the first
BZ consist of one circular and one intersecting-elliptical hole pockets at the BZ center $\Gamma$, and one elliptical electron pocket at the
BZ boundary $X$ with its long-axis along $\Gamma$-$X$. In addition, there is one small circular electron pocket at $X$ (-$\pi$, -$\pi$) in the
second BZ. These experimental FS topology is consistent with the theoretical calculations in Fig. 2(f). Moreover, we observe some additional
FSs around the $X$ points in Figs. 2(b) and 2(d), which could be associated with the lattice periodic potential of the termination layer on the (001)
surface. As the ARPES experimental technique with vacuum ultraviolet lights is surface sensitive, the excited photoelectrons suffer from
the influence of periodic potential on the (001) surface. As the consequence of broken translational symmetry along the (001) direction,
the projected (001) surface BZ, illustrated as dashed lines in Fig. 2(d), is reduced by 1/$\sqrt{2}$ as compared with the translation
periodicity in the $k_x$-$k_y$ plane of bulk BZ. As a result, the band folding with a wave vector $Q$ = ($\pi$, $\pi$, 0) leads to the
additional FSs around $X$. By comparing the FS areas in ARPES with the $A_k$ values in QO, we assign the electron FSs at $X$ (-$\pi$, -$\pi$)
and (-$\pi$, 0) as $\alpha$ and $\gamma$, respectively, and the inner and outer hole FSs at $\Gamma$ as $\beta$ and $\delta$, respectively,
as indicated in Fig. 2(e). With the knowledge of the FS profiles, we derive the FSs from the $A_k$ values and plot them as solid curves in
Fig. 2(e), which are well consistent with those determined by ARPES.

As LaSb has a cubic crystal structure, its electronic structure is identical along the three directions $k_x$, $k_y$, and $k_z$
of the BZ. The 3D structure of the FSs can thus be reconstructed by the regular 2D FSs in the $k_z$ = 0 and $\pi$ planes. The 3D
FSs consist of one spherical and one intersecting-ellipsoidal hole FSs at $\Gamma$ reconstructed by the $\beta$ and $\delta$ FSs,
respectively, and one ellipsoidal electron FS at $X$ reconstructed by the $\alpha$ and $\gamma$ FSs, which is elongated along
$\Gamma$-$X$, as seen in Fig. 2(f). The regular 3D structure of these FSs enables us to precisely quantify the volume of each FS.
Note that while the ARPES measurements do not map the FSs exactly at $k_z$ = 0 and $\pi$, the extreme cross sections of the FSs
at $k_z$ = 0 and $\pi$ can be accurately extracted from the QO data. Therefore, we calculate the volumes of the 3D FSs that are
reconstructed with the 2D FSs derived from the QO data. The enclosed volumes of the $\beta$, $\delta$ and $\alpha$/$\gamma$ FSs
are 0.00639, 0.0196, and 0.00856 $\AA^{-3}$, corresponding to carrier densities of 5.15$\times$10$^{19}$, 1.58$\times$10$^{20}$,
and 6.90$\times$10$^{19}$ cm$^{-3}$, respectively. Considering that there are three electron $\alpha$/$\gamma$ FSs in one BZ, the
ratio of electron-to-hole carrier density $n_e$/$n_h$ = 0.99, indicating LaSb a compensated semimetal within the experimental
accuracy. This is in conflict with the claim in Ref. \cite{18}, in which it is argued that LaSb is not compensated, but our
result is in agreement with the previous reports \cite{42, 43}. Furthermore, the extracted FSs from the ARPES data measured at
$T$ = 30 and 200 K are exactly the same, as shown in Fig. 2(e), indicating that the condition of carrier compensation is held
in a broad temperature range.

Figure 3 shows the near-$E_F$ band dispersions along the high-symmetry lines recorded with $hv$ = 53 eV, whose momentum locations
are indicated in Fig. 2(b). We first discuss the topological nature of LaSb based on the band dispersions. First-principles electronic
structure calculations show controversial results about the band topology, depending on the density functions used in the calculations
\cite{18, 41, 44}. In the calculation with the Perdew-Burke-Ernzerhof (PBE) function at the generalized gradient approximation (GGA)
level, the band inversion happens at $X$ between the La $d$-states and the Sb $p$-states, suggesting that LaSb is a 3D topological
insulator, whereas the calculation with the modified Becke-Johnson (MBJ) potential \cite{45, 46} at the meta-GGA level shows no band
inversion at $X$ \cite{18, 41}. The major discrepancy between the band structures in these two types of calculations is that there
exists an anti-crossing of two bands along $\Gamma$-$X$ in the former. To elucidate the band topology of LaSb, we analyze the
experimental band dispersions along $\Gamma$-$X$. As shown in Figs. 3(a) and 3(b), on moving from $\Gamma$ to $X$, the outer hole
band gradually levels off and then curves upward, forming a hole band with a top at $\sim$-0.35 eV at $X$. In addition, there is a
parabolic electron band along $\Gamma$-$X$ with a bottom at $\sim$-0.25 eV at $X$, forming a band gap of $\sim$0.1 eV. Note that the
linear electron band at $X$ in Figs. 3(a)-3(c), which is associated with the additional FSs in Fig. 2(b), is a result of the band
folding with a wave vector $Q$ = ($\pi$, $\pi$, 0). It is clear that there is no band anti-crossing along $\Gamma$-$X$, which is at
odds with the calculation with the PBE functional. In contrast, the experimental band dispersions are well reproduced by the
calculations with the MBJ potential. We therefore conclude that LaSb is a topologically trivial material.

As our results confirm that LaSb is topologically trivial, no Dirac-like surface states are observed. Therefore, the 2D angle
dependence of QOs reported previously \cite{18} could not be a result of 2D surface state FSs associated with a topologically
nontrivial phase. In contrast, the 2D transport behavior should arise from the ellipsoidal $\alpha$/$\gamma$ FS, which is largely
elongated along $\Gamma$-$X$ with the ratio of long-axis to short-axis $\sim$4. We further reveal that the highly anisotropic
electron band associated with the $\alpha$/$\gamma$ FS disperses parabolically along the long-axis [Figs. 3(a)-3(c)] but linearly
along the short-axis [Figs. 3(f)-3(j)]. Many materials exhibit ultrahigh mobility possess linearly dispersive bands, such as Bi
\cite{47}, Cd$_3$As$_2$ \cite{27}, TaAs \cite{28}, and NbP \cite{29}. It is widely believed that the high mobility in these
materials is associated with the linear bands. The linearly dispersive electron band may be the origin of the high mobility
in LaSb.

We examine the proposed mechanisms for the XMR by comparing the electronic structure of LaSb with those of other XMR semimetals.
Firstly, in WTe$_2$ the measured FSs show dramatic changes with varying temperature \cite{33, 34}, which spoils the carrier
compensation. This was considered as the origin of the drastically suppressed XMR with increasing temperature. The suppression
of MR is also observed in LaSb, for instance the MR at 200 K is suppressed by three orders as compared to that at 2 K under 9 T.
If the suppression is attributed to carrier imbalance, the ratio $n_e$/$n_h$ has to be $\sim$100 in the simulation based on the
two-band model, as plotted in Fig. 1(e). This is contradictory to our ARPES results, which reveal no observable changes of the
FSs with temperature in LaSb. Secondly, due to the lack of inversion symmetry in WTe$_2$, the spin degeneracy is removed by
spin-orbit coupling, leading to a complicated spin texture of the bands, which has been claimed to play an important role in
the XMR of WTe$_2$ \cite{31}. This explanation is not applicable to LaSb because the inversion symmetry in LaSb preserves that
its spins are doubly-degenerate at zero field. Thirdly, the field-induced insulator-like resistivity with a plateau is considered
as the consequences of breaking time reversal symmetry in topological semimetals \cite{18}, whereas our results have demonstrated
that LaSb is a topologically trivial material, which excludes the possibility that the XMR is associated with nontrivial band
topology. Fourthly, it is worth noting that the bulk electronic structure of LaSb shares a considerable degree of similarity
to that of the Weyl semimetal TaAs family, which consists of hole FSs from the normal parabolic bands and electron FSs from
the linear Weyl bands \cite{29, 49}. These Weyl semimetals exhibit a linear MR \cite{28, 29} distinct from the near-quadratic
behavior in LaSb, suggesting they should be associated with different origins.

While LaSb satisfies the two prerequisites for XMR in the two-band model, $i.e.$, carrier compensation and ultrahigh mobility,
the deviation from the quadratic behavior cannot be explained by a slightly imperfect carrier compensation. Thus we consider a
three-band model within the general multiple-band picture \cite{30}. Our ARPES results exhibit that the electron band is highly
anisotropic along and perpendicular to the long-axis of the ellipsoidal FS while the two hole bands have similar dispersions near
$E_F$. Based on the electronic structure, considering the magnetic field applied along the (001) direction, one can construct a
three-band model including two kinds of electrons and one kind of holes. As seen in Fig. 1(d), the formula derived from the
three-band model with carrier compensation \cite{30} fits to the MR curve excellently. Further investigation is desirable to
clarify if the near-quadratic XMR can be understood in the frame of the multiple-band picture. As a common behavior in many
materials, the XMR remains puzzling by the complication from the band structures. LaSb with a simple electronic structure
represents an ideal system to formulate a theoretical understanding to this exotic phenomenon.

\begin{acknowledgments}
This work was supported by the Ministry of Science and Technology of China (Nos. 2012CB921701, 2016YFA0302400, 2016YFA0300600,
2013CB921700 and 2015CB921300), the National Natural Science Foundation of China (Nos. 11274381, 11274362, 11474340, 11274367,
11474330, 11574394 and 11234014), and the Chinese Academy of Sciences (No. XDB07000000). KL was supported by the Fundamental
Research Funds for the Central Universities, and the Research Funds of Renmin University of China (RUC) (Nos. 14XNLQ03, 15XNLF06).
Computational resources was provided by the Physical Laboratory of High Performance Computing at RUC. The FSs were prepared with
the XCRYSDEN program \cite{50}.
\end{acknowledgments}

\end{document}